\documentclass{nature}

\def\xmm{{\it XMM-Newton}}

\def\igr{IGR J18245--2452}
\def\psr{PSR J1824--2452I}
\def\ltsima{$\; \buildrel < \over \sim \;$}
\def\simlt{\lower.5ex\hbox{\ltsima}}
\def\gtsima{$\; \buildrel > \over \sim \;$}
\def\simgt{\lower.5ex\hbox{\gtsima}}

\usepackage{float}
\usepackage{graphicx}

\usepackage{rotate}

\usepackage{lscape}
\usepackage{longtable}

\title{Swings between rotation and accretion power in a millisecond binary pulsar}

\author{A.~Papitto$^{1}$, C.~Ferrigno$^{2}$, E.~Bozzo$^{2}$,
  N.~Rea$^{1}$, L.~Pavan$^{2}$, L.~Burderi$^{3}$, M.~Burgay$^{4}$,
  S.~Campana$^{5}$, T.~Di Salvo$^{6}$, M.~Falanga$^{7}$,
  M.~D.~Filipovi\'c$^{8}$, P.~C.~C.~Freire$^{9}$,
  J.~W.~T.~Hessels$^{10,11}$, A.~Possenti$^{4}$, S.~M.~Ransom$^{12}$,
  A.~Riggio$^{3}$, P.~Romano$^{13}$, J.~M.~Sarkissian$^{14}$,
  I.~H.~Stairs$^{15}$, L.~Stella$^{16}$, D.~F.~Torres$^{1,17}$,
  M.~H.~Wieringa$^{18}$ G.~F.~Wong$^{8,14}$}

\begin{document}

\maketitle

\begin{affiliations}
 \item Institute of Space Sciences (ICE; IEEC-CSIC), Campus UAB,
   Faculty of Science, Torre C5, parell, 2a planta, E-08193 Barcelona,
   Spain
 \item ISDC, Department of Astronomy, 
 Universit\'e de Gen\`eve, chemin d'\'Ecogia, 16, CH-1290
     Versoix, Switzerland
  \item Dipartimento di Fisica, Universit\'a di Cagliari, SP Monserrato-Sestu, Km 0.7, I-09042 Monserrato, Italy
\item INAF–-Osservatorio Astronomico di Cagliari, loc. Poggio dei Pini, strada 54, I-09012 Capoterra, Italy
 \item INAF--Osservatorio Astronomico di Brera, Via Bianchi 46,
    I-23807 Merate, Lecco, Italy
    \item Dipartimento di Fisica e Chimica, Universit\'a di Palermo, via
      Archirafi 36, I-90123 Palermo, Italy
\item International Space Science Institute, Hallerstrasse 6, CH-3012
   Bern, Switzerland
\item University of Western Sydney, Locked Bag 1797, Penrith South DC, NSW 1797, Australia
\item Max-Planck-Institut f{\'u}r Radioastronomie, auf dem H{\'u}gel 69, 53121, Bonn, Germany
\item ASTRON, the Netherlands Institute for Radio Astronomy, Postbus 2, 7990 AA, Dwingeloo, The Netherlands
\item Astronomical Institute ``Anton Pannekoek'', University of Amsterdam, Science Park 904, 1098 XH Amsterdam, The Netherlands
\item National Radio Astronomy Observatory (NRAO), 520 Edgemont Road,
Charlottesville, VA 22901, USA.
  \item INAF-Istituto di Astrofisica Spaziale e Fisica Cosmica, Via U. La Malfa 153, I-90146 Palermo, Italy
\item  CSIRO Astronomy and Space Science, P.O. Box 76, Epping 1710, Australia
\item Department of Physics and Astronomy, University of British
Columbia, 6224 Agricultural Road, Vancouver, British
Columbia V6T 1Z1, Canada
\item INAF-Osservatorio Astronomico di Roma, Via di Frascati 33, I-00040 Monte Porzio Catone, Roma, Italy
\item Instituci\'o Catalana de Recerca i Estudis Avan\c{c}ats (ICREA), 08010 Barcelona, Spain
\item  CSIRO Astronomy and Space Science, Locked Bag 194, Narrabri NSW 2390, Australia

\end{affiliations}

\begin{abstract}

 It is thought that neutron stars in low-mass binary systems can
 accrete matter and angular momentum from the companion star and be
 spun-up to millisecond rotational
 periods\cite{bisnovatyikogan1974,alpar1982,radhakrishnan1982}.
 During the accretion stage, the system is called a low-mass X-ray
 binary and bright X-ray emission is observed. When the rate of mass
 transfer decreases in the later evolutionary stages, these binaries
 host instead a radio millisecond
 pulsar\cite{backer1982,ruderman1989}, whose emission is powered by
 the neutron star's rotating magnetic field\cite{pacini1967}.  This
 scenario is supported by the detection of X-ray millisecond
 pulsations from several accreting neutron
 stars\cite{wijnands1998,chakrabarty1998} and the evidence for a past
 accretion disc in a rotation-powered millisecond
 pulsar\cite{archibald2009}. It has been proposed that a
 rotation-powered pulsar may temporarily switch
 on\cite{stella1994,campana1998,burderi2001} during periods of low
 mass inflow\cite{vanparadijs1996} in some such systems. However, only
 indirect evidence for this transition had been
 observed\cite{burderi2003,hartman2008,disalvo2008,patruno2010,papitto2011}. Here
 we report the detection of accretion-powered, millisecond X-ray
 pulsations from a neutron star previously seen as a rotation-powered
 radio pulsar.  Within a few days following a month-long X-ray
 outburst, radio pulses were again detected.  This not only
 demonstrates the evolutionary link between accretion and
 rotation-powered millisecond pulsars, but also that some systems can
 swing between the two states on very short timescales.

\end{abstract}

%%%%%%%%%%%%%%%%%%%%%%%%%%%%%%%%%%%%%%%%%%%%%%%%%%%%%%%%%%%%%%%%

The X-ray transient {\igr} was first detected by INTEGRAL on 28 March
2013, and is located in the globular cluster M28. The X-ray luminosity
of a $\mbox{few}\times 10^{36}$ erg s$^{-1}$ (0.3–-10 keV), and the
detection by the X-ray Telescope (XRT) on-board Swift of a burst
originated by a thermonuclear explosion at the surface of the compact
object\cite{lewin1993}, firmly classified this source as an accreting
neutron star with a low-mass companion. An observation performed by
XMM-Newton on 4 April 2013 revealed a coherent modulation of its X-ray
emission at a period of $3.93185$ ms (see Fig.~\ref{fig:lcurve}
and~\ref{fig:orbit}). We observed delays of the pulse arrival times
produced by the orbit of the neutron star around a companion star of a
mass $>0.17\; M_{\odot}$, with an orbital period of $11.0$ hours (see
Fig.~\ref{fig:orbit}). The spin and orbital parameters of the source
were further improved by making use of a second XMM-Newton
observation, as well as two observations performed by Swift/XRT (see
Table~\ref{tab}).

Cross-referencing with the known rotation-powered radio pulsars in
M28, we found that {\psr} has ephemerides\cite{begin2006,ATNF} identical to those of {\igr}
(see Table~\ref{tab}).  However, the X-ray pulsations we have observed
from {\igr} are not powered by the rotation of the magnetic field as
for the radio emission of {\psr}. The pulse amplitude was observed to
vary in strong correlation with the X-ray flux, implying that
pulsations came from a source emitting $\approx 10^{36}$ erg s$^{-1}$
in X-rays; this value is larger by more than two orders of magnitudes
than the luminosity shown by the X-ray counterparts of
rotation-powered radio millisecond pulsars\cite{bogdanov2011}, while
it nicely agrees with the X-ray output of accretion-powered
millisecond pulsars\cite{wijnands1998}. The X-ray spectrum of {\igr}
was also typical of this class, and the broad emission line observed
at an energy compatible with the iron K-$\alpha$ transition
($6.4$-–$6.97$ keV) is most easily interpreted in terms of reflection
of hard X-rays by a truncated accretion
disk\cite{papitto2009}. Furthermore, pulsations were detected by
Swift-XRT during the decay of a thermonuclear burst, following a
runaway nuclear burning of light nuclei accreted onto the neutron star
surface. Such bursts are unambiguous indicators that mass accretion is
taking place\cite{lewin1993}, and the oscillations observed in some of
them trace the spin period of the accreting neutron
star\cite{chakrabarty2003}.

We derived a precise position for {\igr} using a Chandra image taken
on 29 April 2013, while the source was fading in X-rays. Analysis of
archival Chandra observations from 2008 indicate that {\igr} already
showed variations of its X-ray luminosity by an order of magnitude, as
shown in Fig.~\ref{fig:chandra}, suggesting it underwent other
episodes of mass accretion in the past few years. This 2008
enhancement of the X-ray emission followed the nearest previous
detection of the radio pulsar, on 13 June 2008, by less than two
months, indicating a very rapid transition from rotation to
accretion-powered activity (see Table 3 in Supplementary Information
for a summary of past observations of the source in the X-ray and
radio band). The Chandra position of {\igr} is compatible with a
variable unpulsed radio source that we have detected with the
Australia Compact Telescope Array on 2013, April 5, with spectral
properties typical of an accreting millisecond pulsar in
outburst\cite{gaensler1999}.

  A combination of serendipitous and target-of-opportunity
  observations with the Green Bank Telescope (GBT), Parkes radio
  telescope, and Westerbork Synthesis Radio Telescope (WSRT) partially
  map the reactivation of {\igr} as the radio pulsar {\psr} (see Table
  3 in Supplementary Information).  No pulsed radio emission was seen
  in any of the three 2013 April observations, compatible with the
  neutron star being in an accretion phase and inactive as a radio
  pulsar.  We caution however, that a non-detection of radio
  pulsations from {\psr} can also be due to eclipsing and
  that the lack of observable radio pulsations does not necessarily
  prove the abscence of an active radio pulsar
  mechanism\cite{begin2006,bogdanov2011}.  Radio pulsations were
  detected in 5 of the 13 observations conducted with GBT, Parkes, and
  WSRT in 2013 May.  These observations demonstrate that the radio
  pulsar mechanism was active no more than a few weeks after the peak
  of the X-ray outburst.

In the last decade, {\igr} has thus shown unambiguous tracers of both
rotation and accretion powered activity, providing conclusive evidence
for the evolutionary link between neutron stars in low mass X-ray
binaries and millisecond radio pulsars. The source swung between
rotation and accretion powered states on few-day to few-month time
scales; this establishes the existence of an evolutionary phase during
which a source can alternate between these two states over a time
scale much shorter than the Gyr-long evolution of these binary
systems, as they are spun-up by mass accretion to millisecond spin
periods\cite{bhattacharya1991}. It is probable that a rotation powered
pulsar switches on also during the X-ray quiescent states of other
accreting millisecond
pulsars\cite{burderi2003,hartman2008,disalvo2008,patruno2010,papitto2011},
even if radio pulsations were not detected\cite{burgay2003}, so
far.

 The short time-scales observed for the transitions between accretion
 and rotation powered states of {\igr} are comparable with those
 typical of X-ray luminosity variations. Like other X-ray transients,
 {\igr} is X-ray bright ($L_X\approx10^{36}$ erg s$^{-1}$) only during
 a few month-long periods called `outbursts', while outside these
 episodes it spends years in an X-ray quiescent state
 ($L_X\simlt10^{32}$ erg s$^{-1}$). These variations are caused by
 swings of the mass in-flow rate onto the neutron
 star\cite{vanparadijs1996}, and our findings strongly suggest that
 this quantity mainly regulates the transitions between accretion and
 rotation powered activity, compatible with earlier
 suggestions\cite{ruderman1989,stella1994,campana1998,burderi2001}. The
 X-ray luminosity of {\igr} during quiescence ($L_{X}\approx10^{32}$
 erg s$^{-1}$) implies that rate of mass accretion was not larger than
 $\dot{M}\simlt10^{-14}$ M$_{\odot}$ yr$^{-1}$, during such a state.
 The presence of radio millisecond pulsations indicates that the
 pulsar magnetosphere kept the plasma beyond the light cylinder radius
 (located at a distance of $\approx$200 km), despite the pressure
 exerted by the mass inflowing from the companion star.  A pulsar
 magnetic field of the order of $10^8$--$10^9$ G is able to satisfy
 this condition, and to explain the quiescent X-ray luminosity in
 terms of the pulsar rotational power (for a typical conversion
 efficiency of about 1$\,\%$).  The irregular disappearance of the
 radio pulses of {\psr} during the rotation powered stage suggests
 that, during that phase, most of the matter that the companion
 transfers towards the neutron star is ejected by the pressure of the
 pulsar wind\cite{fruchter1988,ruderman1989}. A slight increase of the
 mass transfer rate may subsequently push the magnetosphere back
 inside the light cylinder\cite{burderi2001}. After a disk had
 sufficient time to build up, an X-ray outburst is expected to take
 place, as in the case of {\igr} during the observations reported
 here. As the mass accretion rate decreases during the decay of the
 X-ray outburst, the pressure of the magnetosphere is able to, at
 least partially, sweep away the residual matter from the surroundings
 of the neutron star, and a rotation-powered pulsed radio emission can
 reactivate. Our observations prove that such transitions can take
 place in both directions, on a time scale shorter than expected,
 perhaps only a few days.

 The discovery of {\igr}, swinging between rotation and
 accretion-powered emission, represents the most stringent probe of
 the recycling
 scenario\cite{bisnovatyikogan1974,alpar1982,radhakrishnan1982}, and
 the existence of an unstable intermediate phase in the evolution of
 low mass X-ray binaries, offering the unprecedented opportunity to
 study in detail the transitions between these two states.

%\bibliography{nature.bib}

\begin{addendum}
 \item This letter is based on ToO observations made by XMM-Newton,
   Chandra, INTEGRAL, Swift, ATCA, WSRT, GBT, and PKS. We thank the
   respective directors and operation teams for their support. Work
   done in the framework of the grants AYA2012-39303, SGR2009-811, and
   iLINK2011-0303, and with the support of CEA/Irfu, IN2P3/CNRS and
   CNES (France), INAF (Italy), NWO (The Netherlands), and NSERC
   (Canada). A.~Pa. is supported by a Juan de la Cierva Research
   Fellowship.  A. R. acknowledges Sardinia Regional Government for
   financial support (P.O.R. Sardegna ESF 2007-13). D.~F.~T was
   additionally supported by a Friedrich Wilhelm Bessel Award of the
   Alexander von Humboldt Foundation.  L.~P. thanks the Soci\'et\'e
   Acad\'emique de Gen\`eve and the Swiss Society for Astrophysics and
   Astronomy.  Finally, we acknowledge the use of data supplied by the
   UK Swift Science Data Centre at the University of Leicester.
   A.~Pa. thanks S.~Giannetti, D.~Lai, R.~V.~E.~Lovelace,
   M.~M.~Romanova for stimulating discussions, and Wolf~Sol.~Dig. for
   operational support.
\item[Author Contributions] A.~Pa., C.~F. and E.~B. collected and
  analysed XMM-Newton data. A.~Pa. and C.~F. detected the pulsar in
  XMM-Newton data and derived its orbital solution. A.~Pa. discovered
  the equivalence of its parameters with a radio pulsar, the
  thermonuclear burst and the burst oscillations. N.~R. analysed
  Chandra data, detecting the X-ray quiescent counterpart of the
  source and past accretion events. L.~P., M.~H.~W., M.~D.~F. and
  G.~F.~W. analysed ATCA data. E.~B., S.~C., P.~R., A.~Pa. and
  A.~R. analysed Swift data. E.~B. and C.~F. analysed INTEGRAL
  data. J.~W.~T.~H. analysed WSRT data. M.~B. and J.~M.~S. analysed
  PKS data.  J.~W.~T.~H., S.~M.~R., A.~Po. and I.~H.~S. and
  P. C. C. F. analysed GBT data. A.~R. provided valuable software
  tools. A.~Pa., N.~R., and J.~W.~T.~H. wrote the manuscript, with
  significant contribution by all the authors in interpreting the
  results and editing of the manuscript.
 \item[Competing Interests] The authors declare that they have no
competing financial interests.
 \item[Correspondence] Correspondence and requests for materials
should be addressed to A.~Papitto~(email: papitto@ice.csic.es).
\end{addendum}

%%
%% TABLES
%%
%% If there are any tables, put them here.
%%

\begin{table*}
\footnotesize
\caption{Spin and orbital parameters of {\igr} and {\psr}.}
\label{tab}
\begin{center}
\begin{tabular}{@{}l c c }
\hline \hline
Parameter  & {\igr} & {\psr} \\
\hline
Right Ascension (J2000) &  $18^{h}\;24^{m}\;32.53(4)^{s}$ & \\ 
Declination (J2000) & $-24^{\circ}\;52'\;08.6(6)''$ & \\
Reference epoch (MJD) & $56386.0$ & \\
Spin period (ms) & $3.931852642(2)$ & $3.93185(1)$ \\
Spin period derivative  &  $< 1.3\times10^{-17}$ & \\
RMS of pulse time delays (ms) & $0.1$ & \\
Orbital period (hr) & $11.025781(2)$ & $11.0258(2)$ \\
Projected semi-major axis  (lt-s) & $0.76591(1)$ & $0.7658(1)$ \\
Epoch of zero mean anomaly (MJD) & $56395.216893(1)$ & \\
Eccentricity & $\leq 1\times10^{-4}$ & \\
Pulsar mass function (M$_{\odot}$) & $2.2831(1)\times10^{-3}$ & $2.282(1)\times10^{-3}$ \\
Minimum companion mass (M$_{\odot}$) & $0.174(3)$ & $0.17(1)$ \\
Median companion mass (M$_{\odot}$) & $0.204(3)$ & $0.20(1)$ \\
\hline
\end{tabular}
\end{center}
\footnotesize {\bf Coordinates, spin, and orbital parameters of
  {\igr}={\psr}}. Celestial coordinates of {\igr} are derived from a
Chandra X-ray observation performed using the High Resolution Camera
(HRC-S) on 29 April 2013 (see Fig.~\ref{fig:chandra}). The spin and
orbital parameters of {\igr} were derived by modelling the pulse
arrival time delays of the fundamental frequency component, as
observed in the 0.5--10 keV energy band by the EPIC pn camera on-board
XMM-Newton, and by the X-ray Telescope on-board Swift (see
Fig.~\ref{fig:orbit} and Supplementary Information for details). The
solution covers the interval between 30 March and 13 April 2013. The
peak-to-peak amplitude of the fundamental varied in correlation with
the observed count rate (Spearman's rank correlation coefficient of
$\rho=0.79$ for 45 points, which has a probability of less than
$10^{-10}$ if the variables are uncorrelated) with a maximum of
18$\,\%$. When detected, the second harmonic has an amplitude between
2 and 3$\,\%$.  The minimum and median mass of the companion star were
evaluated for a 1.35 M$_{\odot}$ mass of the neutron star, and
inclination of the system of $90^{\circ}$ and $60^{\circ}$,
respectively. The spin and orbital parameters of {\psr} were taken
from ref.~20 and the ATNF pulsar Catalogue\cite{ATNF}, considering
errors on the last significant digit there quoted. The numbers in
parentheses represent the uncertainties on the respective parameter
evaluated at a 1$\sigma$ confidence level. Upper limits are quoted at
a 3$\sigma$ confidence level.

\end{table*}

\begin{figure}
\centering
\includegraphics[width=10cm]{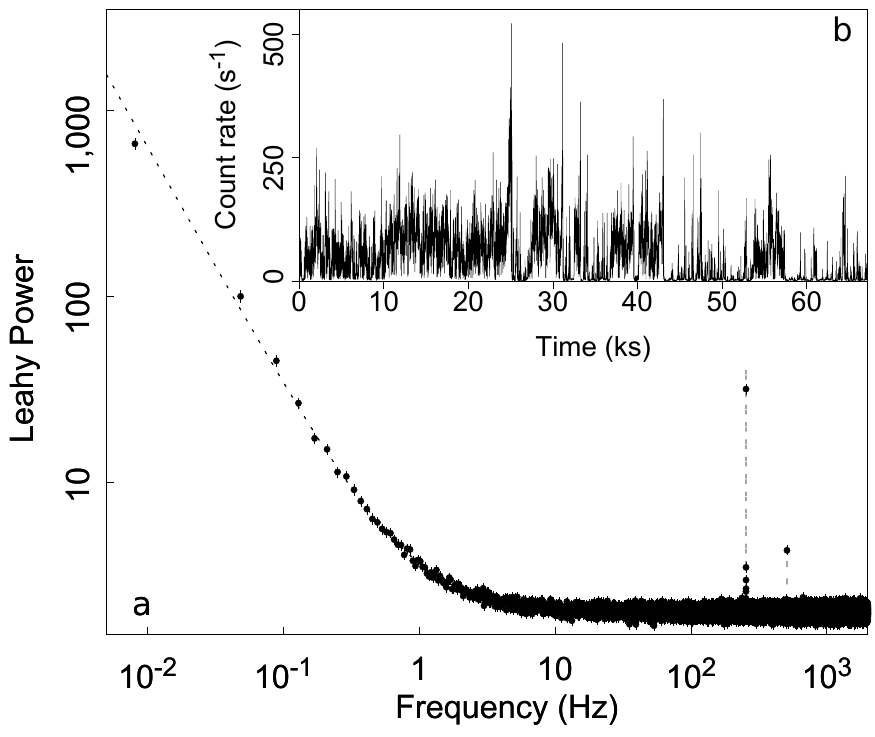}
\caption{\footnotesize {\bf Variability of the X-ray emission of
    {\igr}}. {\bf a}, Fourier power spectral density of the 0.5--10
  keV X-ray photons observed by the EPIC pn camera on-board
  XMM-Newton, during an observation starting on 13 April 2013, for an
  exposure of $67.2$ ks (Obs.~ID 0701981501). The power spectrum was
  obtained by sampling the light curve with a time binning of
  $0.236\;\mbox{ms}$, and averaging intervals of 128 s of length. The
  times of arrival of photons were converted to the barycentre of the
  Solar System and to the line of nodes of the binary system hosting
  {\igr}, by using the parameters listed in Table~\ref{tab}.  The
  peaks at 254.3 and 508.6 Hz represent the first and second harmonic
  of the coherent modulation of the X-ray emission of
  {\igr}. Considering photons observed during a 2-ks interval, not
  corrected for the pulsar orbital motion, the signal at the spin
  period of the neutron star is detected at a significance $\simgt
  80\sigma$. The dashed solid line is the sum of a power-law noise
  function, $P(f)\propto f^{-\gamma}$, with $\gamma=1.291(4)$, and of
  a white noise spectrum with an average value of $1.9900(2)$
  Hz$^{-1}$. Even considering the whole length of the time series, no
  break of the power-law noise could be detected at low
  frequencies. {\bf b}, 0.5--10 keV light curve of the same
  observation, with a bin time of $5$ s.  The possibility of
  contamination by soft proton flares was ruled out by extracting a
  light curve from a background region observed by EPIC MOS cameras
  far from the source. Similar properties of variability than those
  shown here are observed during an XMM-Newton observation starting on
  3 April 2013, for an exposure of $26.7$ ks (Obs.~ID
  0701981401).\label{fig:lcurve}}
\end{figure}

\begin{figure}
\centering
\includegraphics[width=10cm]{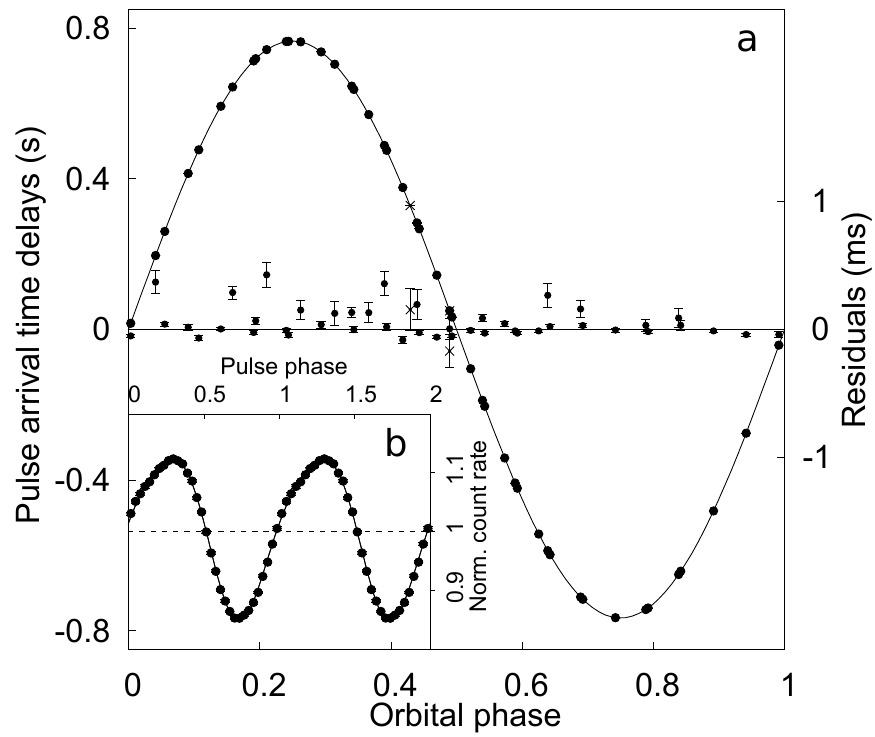}
\caption{\footnotesize {\bf Spin and orbit of {\igr}}. {\bf a}, Pulse
  arrival time delays caused by the orbital motion of the neutron star
  (left axis) as measured by XMM-Newton during observations starting
  on 3 and 13 April 2013 (dots; Obs.~ID 0701981401 and 0701981501,
  with exposure of 26.7 and 67.2 ks , respectively), and by Swift
  during observations starting on 30 March and 7 April 2013 (crosses;
  Obs.~ID 00552369000 and 00032785005, with exposure of 0.6 and 1.6
  ks, respectively). Residuals with respect to best fit timing
  solution (solid line) are also shown (right axis). Pulse profiles
  observed in 2~ks long intervals were modelled using $n = 12$ phase
  bins. The significance of each detection was assessed from the
  probability that the variance of each folded pulse profile were
  compatible with counting noise, assuming that in absence of any
  signal the latter is distributed as a chi-squared variable with
  $(n-1)$ degrees of freedom\cite{leahy1983}. Only detections with a
  significance larger than $3\sigma$ were considered. Pulse arrival
  time delays were determined through standard methods of least square
  fitting of the pulse profiles\cite{papitto2009}, using two harmonic
  components and considering the values measured for the fundamental
  frequency component.  {\bf b}, Average pulse profile sampled in 32
  phase bins, accumulated over the two XMM-Newton observations (black
  dots), and the best fit decomposition with two harmonics (solid
  line). The amplitude of the first and second harmonic was 13.4(1)
  and 1.9(1)$\,\%$, respectively. Two cycles are plotted for
  clarity. In both panels, plotted error bars are the standard
  deviation of each measure.
 \label{fig:orbit}}
\end{figure}

\begin{figure}[H]
\centering
\includegraphics[width=10cm]{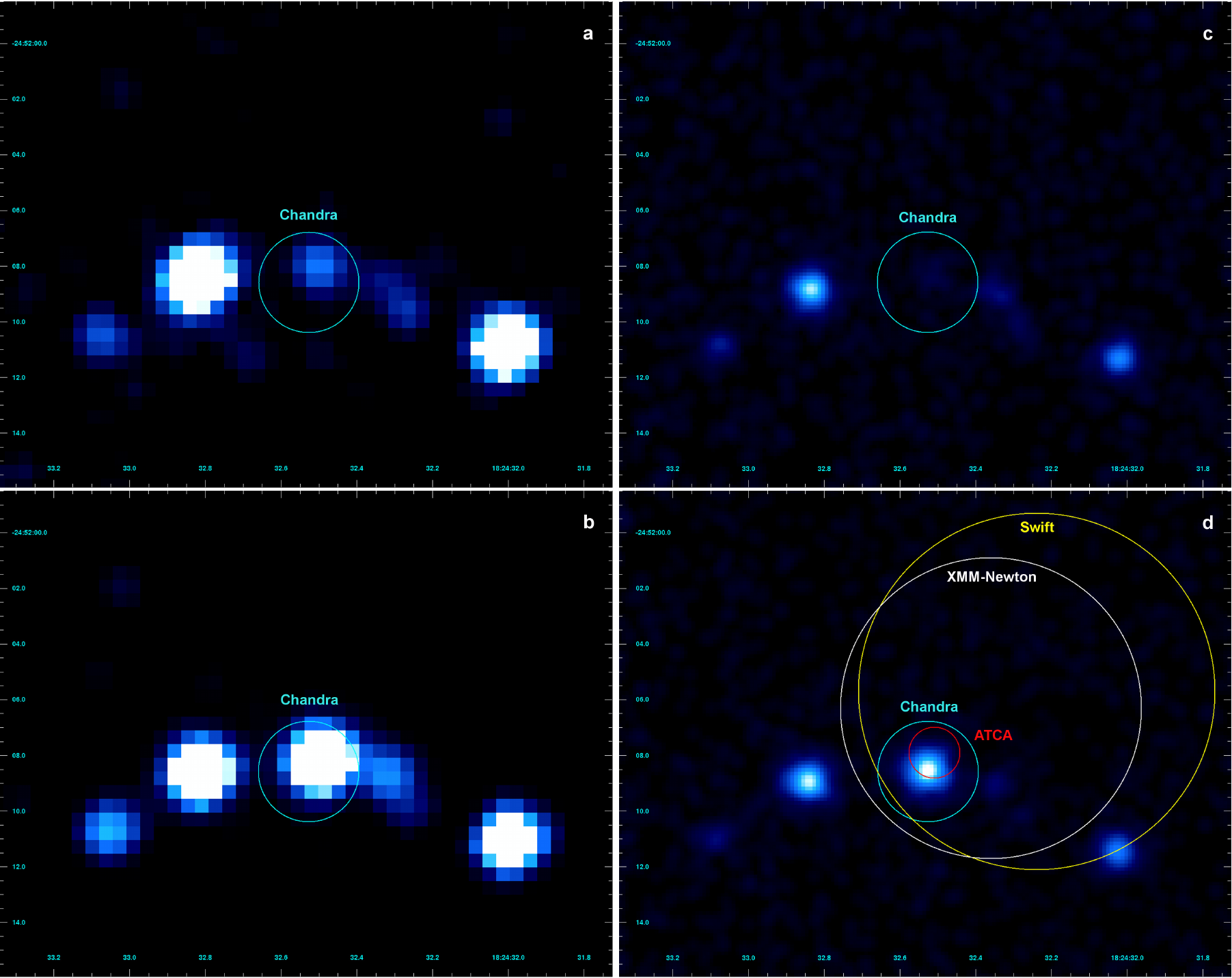}
\centering
\caption{\footnotesize {\bf Long term X-ray variability of
    {\igr}}. {\bf a, b}, Chandra/ACIS-S images of the core of M28
  taken during 4 August 2002, and 7 August 2008, respectively. {\bf c,
    d}, Chandra/HRC-S images of the same field taken during 27 May
  2006, and 29 April 2013, respectively. Images in the upper panels
  ({\bf a, c}) shows the source in X-ray quiescence, emitting a
  luminosity of few$\;\times10^{32}$ erg s$^{-1}$, while during
  observations of the lower panels ({\bf b, d}) the X-ray luminosity
  was of few$\;\times10^{33}$ erg s$^{-1}$ (see Supplementary
  Information for details). The luminosity emitted during the 2013
  observation ({\bf d}) was three orders of magnitude lower than that
  observed by Swift ($3.5\times10^{36}$ erg s$^{-1}$ on 30 March 2013)
  and XMM-Newton ($1.1\times10^{36}$ erg s$^{-1}$ on 3 April 2013) at
  the onset of the X-ray outburst, compatible with the source being
  close to the end of the accretion episode. A
  distance\cite{harris1996} of 5.5 kpc was considered to derive these
  estimates. During the 2013 outburst, the 0.5-10 keV spectrum of
  {\igr} is dominated by a $\approx1.4$ power law interpreted as
  Comptonization in an optically thin medium, of seed photons with a
  temperature of $\approx0.3$ keV. XMM-Newton observations also
  detected a thermal component, modelled as an accretion disc
  truncated at an apparent projected inner radius of $\approx 50$ km,
  and a broad line, modelled with a Gaussian centred at $6.74\pm0.11$
  keV and of $1.1\pm0.2$ keV of width, compatible with iron K-$\alpha$
  transition (see Supplementary Information for details). The plotted
  error circles represent the 3$\sigma$ confidence level position of
  {\igr}, derived by Chandra (29 April 2013), XMM-Newton EPIC-MOS (3
  and 13 April 2013), Swift XRT (30 March 2013), and ATCA (5 April
  2013), plotted as a cyan, white, yellow and red circles,
  respectively. \label{fig:chandra}}
\end{figure}

\title{\bf \Large Supplementary Information}
\maketitle

\section{INTEGRAL detection of {\igr}}

\begin{figure}[h!]
\centering
\includegraphics[width=14cm]{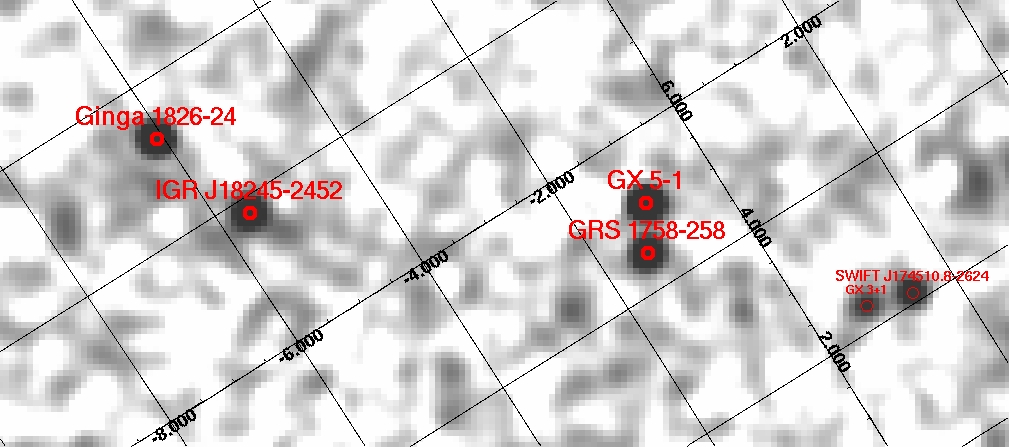}
\caption{\footnotesize Mosaic of the IBIS/ISGRI field of view around
{\igr} obtained by using Science Windows 83-107 collected in the
direction of the source during satellite revolution 1276 (20--40 keV
energy range).\vspace{1.0cm}}
\label{fig:integral}
\end{figure}

{\igr} was first detected\cite{eckert2013} by the hard X-ray imager
IBIS\cite{ubertini03} using the ISGRI\cite{lebrun03} detector on-board
INTEGRAL\cite{winkler2003}, on 28 March 2013, during observations of
the Galactic Center. We analysed the corresponding data (pointings
from 83 to 107 in satellite revolution 1276) by using the Off-line
Science Analysis software provided by the Integral Science Data
Centre. The source was detected in the ISGRI mosaicked image at a
significance level of 21$\sigma$ in the 20--40 keV energy band and
15$\sigma$ in the 40--80 keV energy band (see
Figure~\ref{fig:integral}). The best source position was obtained at
RA=276.14$^{\circ}$, Dec=−24.88$^{\circ}$ (J2000), with an associated
uncertainty of 1.4$'$ at 90$\%$ confidence level, well within the
globular cluster M28. During the first detection of the source, its
flux estimated from the ISGRI data was $9\times10^{-10}$ erg cm$^{-2}$
s$^{-1}$ in the 20-100 keV energy band, during an effective exposure
time of 32 ks, corresponding to a luminosity of $3\times10^{36}$ erg
s$^{-1}$, at a distance\cite{harris1996} of 5.5 kpc. The source was
outside the field of view of the JEM-X instrument\cite{lund03},
sensitive in the 3--30 keV energy band, during the entire observation.

\section{XMM-Newton observations of {\igr}}
\label{sec:xmm}
Following the detection of {\igr} by INTEGRAL, we obtained two target
of opportunity observations with the X-ray Multi-Mirror
Mission\cite{jansen2001} (\xmm), starting on 3 April 2013, at 23:49
(Obs.~ID 0701981401, OBS~1, hereafter) and 13 April at 06:25 (Obs.~ID
0701981501, OBS~2), and lasting 26.7 and 67.2 ks, respectively (all
the epochs are given in the Coordinated Universal Time). The European
Photon Imaging Camera (EPIC) pn\cite{struder2001} was operated in fast
timing mode, the two EPIC MOS cameras\cite{turner2001} in small window
imaging mode, and the Refection Grating
Spectrometers\cite{denherder2001} (RGS) in standard spectroscopy
mode. A thick blocking filter was used to shield the EPIC cameras from
contamination of optical light.  All the data were reduced by using
the latest version of the XMM-Newton Science Analysis Software (SAS
ver. 13). The average count rates observed by the EPIC pn, EPIC
MOS1, EPIC MOS2, RGS1 and RGS2 (in their first order of dispersion)
during the first (second) observations were 68.1 (71.0), 11.8 (11.7),
11.6 (11.5), 0.68 (0.86) and 0.69 (0.88) s$^{-1}$, respectively.
Since the source is detected also at energy larger than 10 keV, we
built a 0.5--10 keV light curve from a background region falling in
one of the outer chips of the EPIC MOS cameras which was not pointing
the source, in order to ensure that soft proton flares of Solar origin
did not contaminate the observations.

\subsection{Timing Analysis.}
\label{sec:temporal}

The timing observing mode of the EPIC pn camera has a timing
resolution of 29.52 $\mu$s. This resolution is achieved by losing
spatial information along one of the axes of the CCD. To extract the
source emission, we considered X-ray photons falling within 86.1$''$
from the source position measured along one of the pointing axes
(corresponding to a full width of 21 EPIC pn pixels), and containing
95$\%$ of the energy at every observed wavelength. The background
emission was extracted from a region of width 12$''$ (corresponding to
3 EPIC pn pixels).

Although X-ray pulsations from several accreting millisecond pulsars
have already been observed by
XMM-Newton\cite{gierlinski2005,papitto2009,patruno2009,papitto2010,papitto2013},
this is the first time that pulsations from an accreting millisecond
pulsar have been discovered by this observatory.  In order to perform
a timing analysis of the signal, the times of arrival of X-ray photons
were converted to the Solar System barycentre, by using the position
determined by Chandra (see Sec.~\ref{sec:chandra} below) and Solar
System ephemerides JPL DE405. We derived a zeroth order orbital solution
by measuring the spin period in 2-ks long intervals through an epoch
folding technique\cite{leahy1983}, and modelling the orbital
modulation affecting the values obtained.  To assess the significance
of any detection we considered that in the absence of any periodic
signal, the variance of a folded profile follows a $\chi^2_{n-1}$
distribution with $n-1$ degrees of freedom, where $n$ is the number of
phase bins used to sample the signal period\cite{leahy1983} ($n=12$ in
this case).  We then used the preliminary determination of the pulsar
orbital parameters to convert the photon arrival times to the line of
nodes of the binary system. This procedure was iterated until the spin
period was observed to be constant throughout the observations. To
further refine the parameters, we then folded data around the current
best estimate of the spin period, considering 12 phase bins and
describing the pulse profile with two harmonic components. The
variation of the phase of the first harmonic over time was modelled in
terms of the difference between the orbital and spin parameters used
to correct the photon arrival times, and the actual
ones\cite{deeter1981}. The procedure was iterated until no significant
corrections to the parameters were found within the
uncertainties\cite{papitto2011}.  We checked that the phase difference
between the first and second harmonic component was compatible with a
constant.

\begin{figure}[h!]
\centering
\includegraphics[width=10cm]{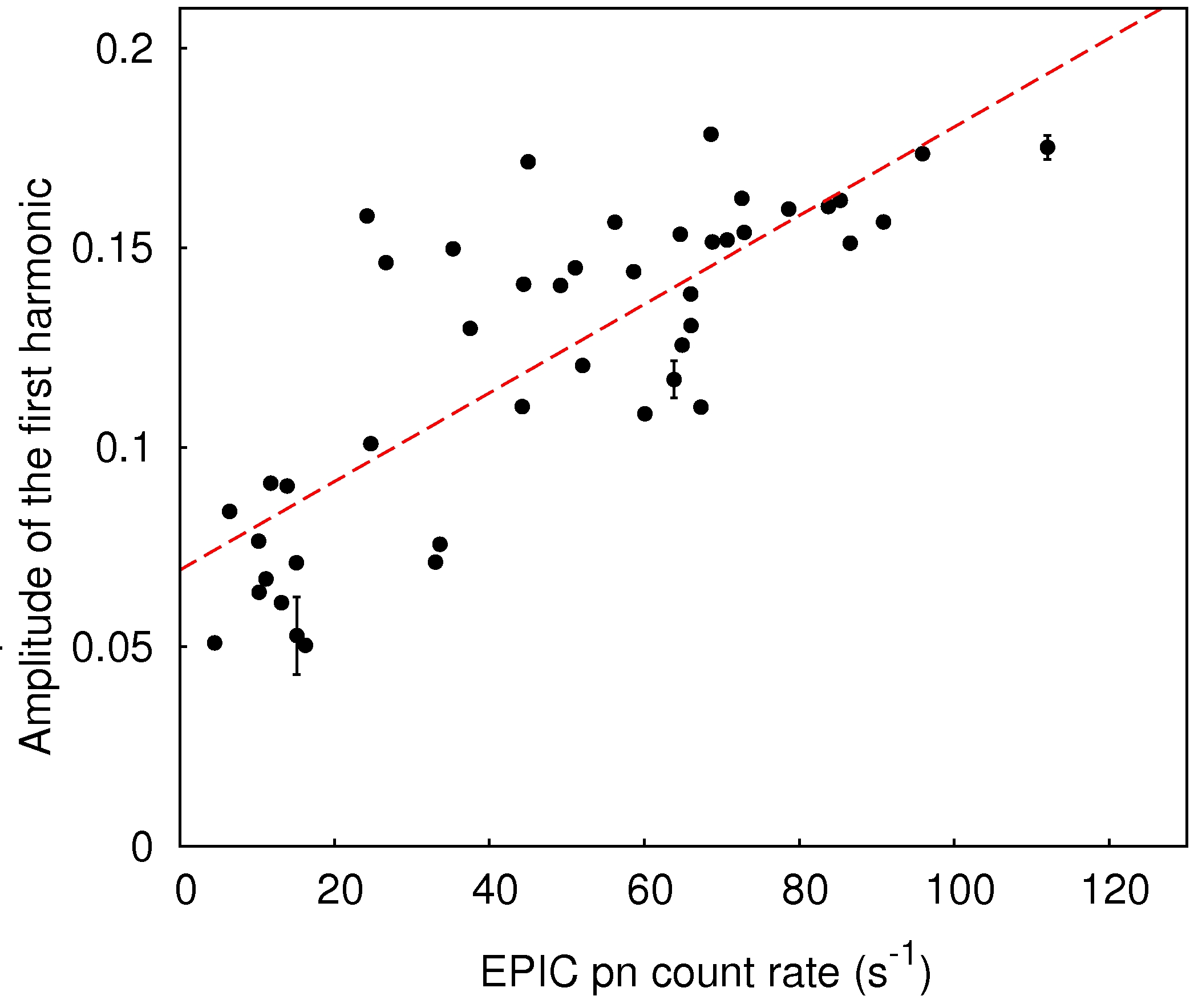}
\caption{\footnotesize Peak-to-peak amplitude of the first
harmonic measured during XMM-Newton OBS~1 and OBS~2, normalised to the
average count rate, and 0.5-10 keV EPIC pn count rate evaluated over
$2$ ks long intervals. The dashed red line is the best fit linear
regression between the two variables. Typical errors bars are shown
for sample points.\vspace{1cm}}
\label{fig:corr}
\end{figure}

The amplitude of the harmonic component at the fundamental frequencies
varied between 18$\%$ and the non detection, in strong positive
correlation with the count rate observed in the 0.5--10 keV (see
Figure~\ref{fig:corr}). The Spearman's rank correlation coefficient
evaluated from the observed count rate and amplitude is $\rho=0.79$
for 45 points, implying a probability of less than $10^{-10}$ that the
two variables are not correlated.

\subsection{Spectral Analysis.}
\label{sec:spectra}

\begin{figure}[h!]
\centering
\includegraphics[width=10cm]{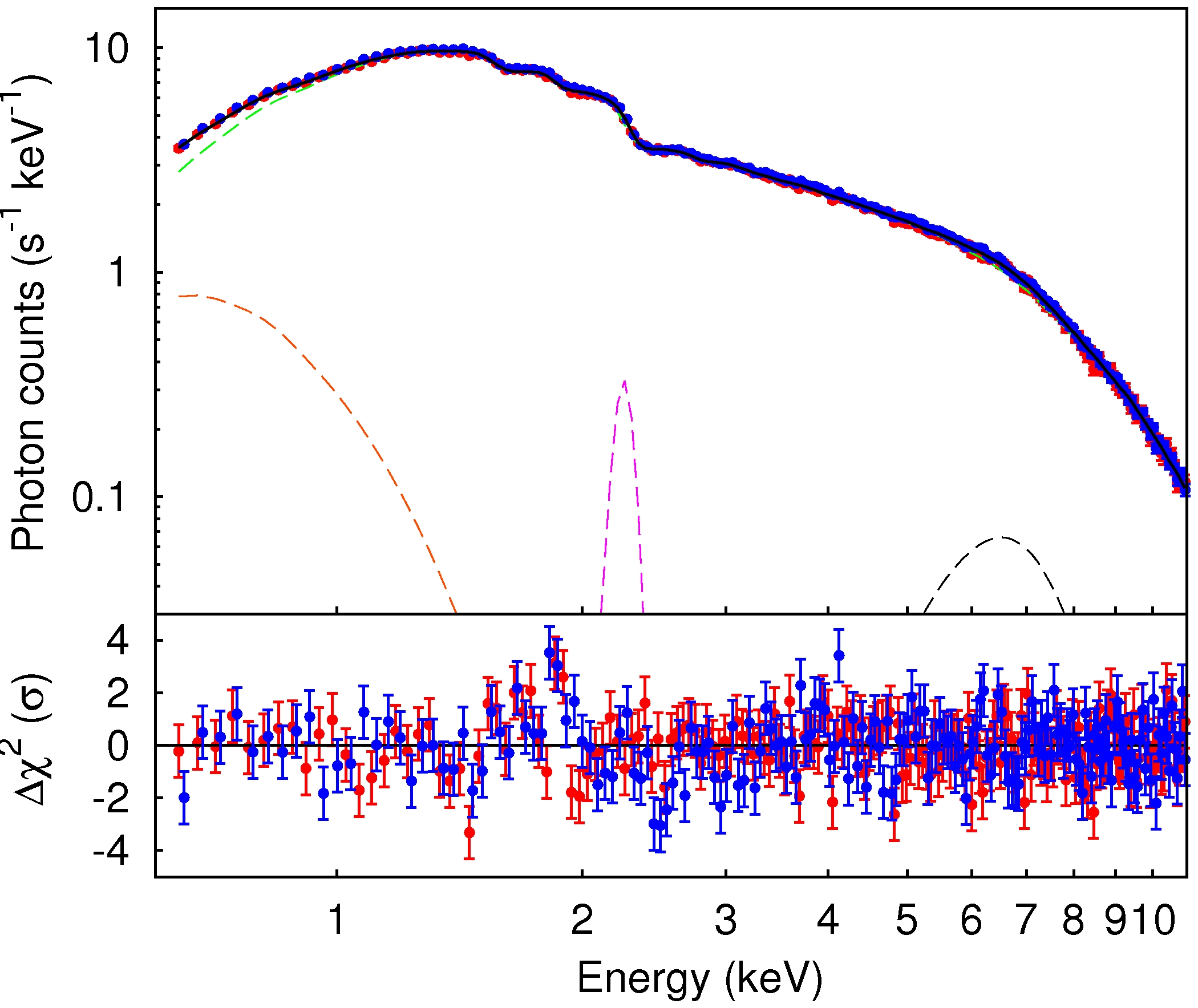}
\caption{\footnotesize X-ray spectrum observed by the EPIC pn
camera during OBS~1 and OBS~2 (red and blue points, respectively). The
best fit model is plotted as a black solid line, the disc emission,
the Comptonized emission, the iron emission feature, and the feature
of calibration origin, are plotted as an orange, green, grey and
magenta dashed lines, respectively (top panel). Residuals in units of
$\sigma$ of the observed spectra, with respect to the best fit model
(bottom panel).\vspace{1.0cm}}
\label{fig:xmmsp}
\end{figure}

We built an X-ray spectrum of the emission observed by the EPIC pn by
excluding the two brightest columns of CCD pixels (RAWX=37--38), to
avoid the effect of photon pile up during the time intervals in which
the count rate was the highest. Spectral bins were grouped to
over-sample the effective instrument resolution by a factor not larger
than three. Response matrices were built following the guidelines
provided by the XMM-Newton calibration technical notes. The two
observations were modelled simultaneously as they had a compatible
average spectral shape. We modelled the 0.6--11 keV observed spectra
with an accretion disk spectral component\cite{makishima1986}
(\texttt{diskbb} in the terminology used by the spectral fitting
routine used, Heasarc's XSPEC v.12.8.0), and with the emission
produced by a thermal distributions of electrons which Compton
up-scatter soft seed X-ray photons, with a black-body spectral
shape\cite{zdziarksi1996,zycki1999} (\texttt{nthcomp}). A similar
spectral decomposition well described the X-ray spectra shown by
accreting millisecond pulsars\cite{gierlinski2002}, observed by the
EPIC pn
camera\cite{gierlinski2005,papitto2009,patruno2009,papitto2010,papitto2013}. The
temperature of the electron cloud was fixed at $50$ keV, outside the
energy band observed by the EPIC pn camera, following the values
usually observed at higher energies from accreting millisecond
pulsars\cite{gierlinski2002,falanga2005} ($kT_e\simgt20$ keV).
Absorption of the interstellar medium was modelled according to the
Tuebingen-Boulder model (Wilms, J. et al. 2011, in preparation,
{http:$//$pulsar.sternwarte.uni-erlangen.de$/$wilms$/$research$/$tbabs$/$}). The
value of the absorption column was fixed to the value measured by
fitting the average spectra observed by the RGS 1 and 2 with an
absorbed power law ($N_{\rm H}=0.31(1)\times10^{22}$ cm$^{-2}$); solar
abundances were considered. A broad Gaussian emission feature centred
at an energy compatible with iron K-$\alpha$ emission (6.4--6.97 keV),
and a narrow emission feature most probably due to calibration
residuals around the edge of Au (Guainazzi, M. et al. 2012, available
from
{http:$//$xmm2.esac.esa.int$/$docs$/$documents$/$CAL-TN-0083.pdf}),
decreased significantly the variance of model residuals. The best-fit
parameters obtained modelling the spectra observed during the two
XMM-Newton are given in Table~\ref{tab:spec}, while the observed
spectrum and residuals are plotted in Figure~\ref{fig:xmmsp}.

\vspace{2cm}

\begin{table*}
\caption{Average spectral parameters of {\igr}.}
\begin{center}
\begin{tabular}{@{}l r}
\hline \hline
Absorption column ($N_{\rm H}$)$^{(a)}$ & $0.31(1)\times10^{22}$ cm$^{-2}$ \\
Inner disk temperature ($kT_{\rm in})$ & $0.13(1)$ keV \\
Apparent inner disk radius ($R_{\rm in}\sqrt{\cos{i}}$)$^{(b)}$ & $52^{+19}_{-13}$ km \\
Seed photon temperature  ($kT_{\rm soft}$) & $0.31(1)$ keV \\
Asymptotic power-law photon index ($\Gamma$) & $1.41(1)$ \\
Electron temperature ($kT_{\rm e}$) & $50$ keV\\
Energy of iron line ($E_{\rm Fe}$) & $6.74\pm0.11$ keV \\
Width  ($\sigma_{\rm Fe}$) & $1.1(2)$ keV \\
Normalisation  ($N_{\rm Fe}$) & $4.8(1)\times10^{-3}$ cm$^{-2}$ s$^{-1}$ \\
Energy of line of calibration origin ($E_{\rm cal}$) & $2.24(1)$ keV \\
Normalisation  ($N_{\rm cal}$) & $1.3(2)\times10^{-4}$ cm$^{-2}$ s$^{-1}$ \\
Unabsorbed flux  (0.5--10 keV; $F_{1}$)$^{(c)}$ & $3.08(1)\times10^{-10}$  erg cm$^{-2}$ s$^{-1}$\\
Unabsorbed flux  (0.5--10 keV; $F_{2}$)$^{(d)}$ & $3.17(1)\times10^{-10}$  erg cm$^{-2}$ s$^{-1}$\\
%Unabsorbed flux of second observation (0.5--10 keV; $F_{1}$) & $3.52(1)\times10^{-10}$ erg cm$^{-2}$ s$^{-1}$\\
\hline
Reduced chi-squared, $\chi_\nu^2 (\mbox{d.o.f.})$ & $1.27 (355)$ \\
\hline
\end{tabular}
\end{center}
%\caption{Parameters based on spectral modelling of the two
\footnotesize Parameters based on spectral modelling of the two
observations of {\igr} performed by the EPIC pn camera. Values in
parentheses are the uncertainties on the last significant digit,
evaluated at a 90$\%$ confidence level.\newline $^{(a)}$ The value of
the interstellar absorption column density was measured by modelling
the spectra observed by the two units of the RGS with an absorbed
power law.  This value was held fixed when fitting the EPIC pn
spectra.  We used abundances and photoelectric cross-sections from
ref.~54 and 55, respectively.\newline $^{(b)}$
The disk apparent radius depends on the disk inclination, $i$, and was
evaluated for a distance of $5.5$ kpc (ref.~35). \newline
$^{(c-d)}$ Unabsorbed flux during OBS~1 and OBS~2,
respectively. \label{tab:spec}
\end{table*}

\subsection{The position.}
\label{sec:mos}

To determine the source position we accumulated images from the EPIC
MOS1 and MOS 2 in the energy bands 0.5-1 keV, 1-2 keV, 2-4.5 keV, and
4.5-12 keV, during the time intervals in which the source count rate
was $<1.5$ s$^{-1}$ (0.5--10 keV). This selection prevented pile-up
and resulted in a total effective exposure time of 5.7 and 13.8 ks for
the two instruments in OBS~1 and OBS~2, respectively. We created an
exposure map using the attitude information to mask out the regions of
the detector where not enough exposure was available, and performed a
first localisation of the sources in the raw images using the sliding
box method\cite{watson2009}. A background map was then produced from
the previous images by removing the identified sources (taking into
account the local instrument point spread function), and used together
with previous products to perform a second optimised source
localisation. The data of the two MOS cameras in the two observations
and different energy bands provided a total of 16 independent
estimates of the source position. By averaging such estimates we
obtained a position of RA=$18^{h}24^{m}32.36^{s}$ Dec=$-24^{\circ} 52'
06.3''$.  We added in quadrature the $1\sigma$ uncertainty we derived
with the described analysis (1.5$''$) to the systematic error
reported in the XMM catalogue\cite{watson2009} (1.0$''$), to obtain
a total uncertainty of 1.8$''$ (1$\sigma$ confidence
level). The error circle at a $3\sigma$ confidence level is plotted as
a white circle in Fig.~3 of the main body of the Letter.

%%%%%%%%%%%%%%%%%%%%%%%%%%%%%%%%%%%%%%%%%%%%%%%%

\section{Swift observations of {\igr}}
\label{sec:swift}

After the initial arc-second localisation with the X-ray
Telescope\cite{burrows2005} (XRT) on-board Swift\cite{gehrels2004},
the source triggered the Burst Alert Telescope\cite{barthelmy2005}
(BAT) on 30 April 2013 at 02:22:21 UT and subsequently at 15:10:37 UT
and 15:17:33.61 UT.  Several observations with XRT were obtained
starting from 30 April, including a intensive one between the BAT
triggers (Obs. Id 32787) and a long term one (Obs. Id 32785).
Analysis of the XRT observation performed on 30 April yielded a
determination\cite{romano2013} of the position of the source
RA=$18^{h}24^{m}32.24^{s}$ Dec=$-24^{\circ} 52' 05.7''$, with an
uncertainty of 3.5$''$ at a 90$\%$ confidence level, compatible with
that determined by ref.~59\nocite{heinke2013}, and those measured by
EPIC MOS (see Sec.\ref{sec:mos} above), Chandra (see
Sec.~\ref{sec:chandra} below), and ATCA (see Sec.~\ref{sec:atca}). The
0.5--10 keV flux corrected for absorption attained a maximum level of
$9.6(1)\times10^{-10}$ erg cm$^{-2}$ s$^{-1}$ during an observation
starting on 30 March 2013 (Obs.ID~00552369000), with a spectral
distribution well described by a power law with index $\Gamma=1.35\pm
0.04$. The last detection was obtained during a 0.8-ks observation
performed on 1 May 2005 (Obs.~ID 00032785021), with an observed flux
of $(1.6\pm0.4)\times 10^{-12}$ erg cm$^{-2}$ s$^{-1}$. This value of
the flux was estimated assuming a $\Gamma=2.5\pm0.6$, power-law shaped
spectrum, as derived summing all Swift observations taken in the week
starting from 21 April 2013. We note that a flux contamination of
$\approx 3\times10^{-13}$ erg cm$^{-2}$ s$^{-1}$ is expected to be
produced by other unresolved source in M28, among which PSR B1821-24
gives the largest contribution\cite{becker2003, bogdanov2011}. Details
of a sample of Swift/XRT observations are given in Supplementary Table
\ref{tab:history}.

\begin{figure}[t!]
\centering
\includegraphics[width=12cm]{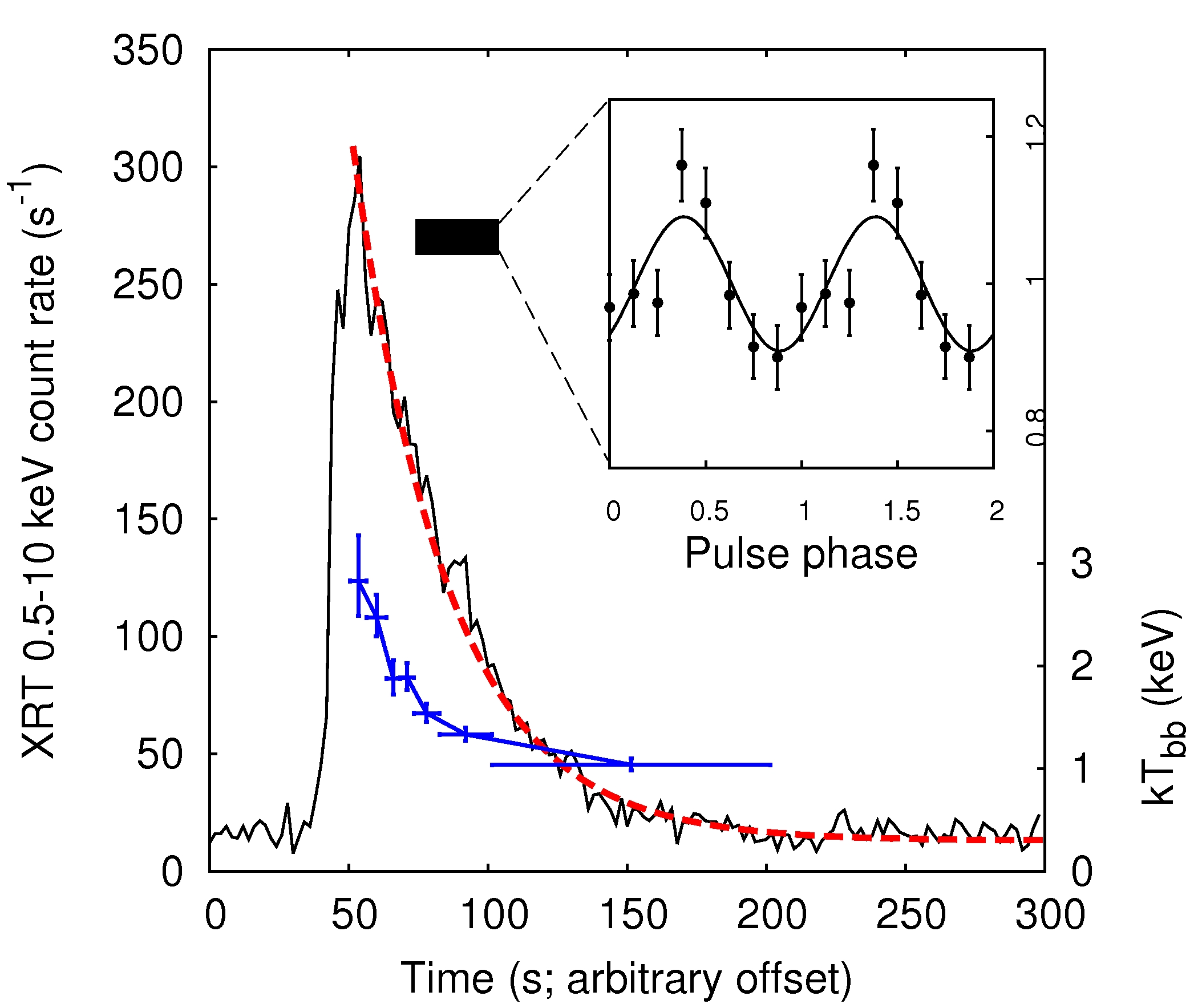}
\caption{\footnotesize Thermonuclear burst observed by XRT, with the best fit
  exponential decay model plotted as a red dashed line. Blue points
  show the temperature of the black body which we used to model the
  XRT 0.5-10 keV spectra, and refer to the right axis. The inset shows
  the pulse profile observed at the spin period of the source, during
  the interval marked by the horizontal thick bar. \vspace{1cm}}
\label{fig:burst}
\end{figure}

During an observation that started on 7 April at 20:32
(Obs.~ID~00032785005), a bursting event was
observed\cite{papitto2013,linares2013} by the XRT when it was
observing in windowed timing mode with a temporal resolution of 1.78
ms. The burst profile has the typical shape observed from
thermonuclear explosions taking place at the surface of the neutron
star\cite{lewin1993}, with a fast rise of $\simlt 10$ s and an
exponential decay on a time scale of $38.9\pm0.5$ s (see red dashed
line in Figure~\ref{fig:burst}). We performed a time resolved analysis
of the emission observed by XRT during the burst. We fit all spectra
with a black body emission absorbed by the interstellar medium, fixing
the value of the absorption column to that indicated by the XMM-Newton
analysis (see Tab.~\ref{tab:spec}). We used the emission observed by
XRT during a 100-s interval before the burst as a background to the
burst emission. The values of temperature we observed are plotted as
blue dots in Figure~\ref{fig:burst}. The temperature decay observed
during the burst tail agrees with the cooling expected after the
thermonuclear burning\cite{lewin1993}, confirming the nature of the
observed event.

We used the Chandra pulsar position (see Sec.~\ref{sec:chandra},
below) and orbital parameters determined from  the XMM-Newton
observations (see Sec.\ref{sec:xmm} and Table 1 in the main body of
the Letter) to report the photons arrival times to the Solar system
Barycentre and to the pulsar line of nodes. We divided the first 160 s
since the burst onset into four intervals, finding in the second one a
significant signal at the pulsar period, with an amplitude of
$13\pm2\%$ (see the inset of Figure ~\ref{fig:burst}). After taking
into account the number of trials made, the detection is significant
at a $3.2\sigma$ level. A signal with an amplitude of $6.0\pm1.2\%$ is
detected during the 740 s observed by XRT before the burst onset.  The
increase of the signal amplitude indicates that the signal observed
during the burst is related to this event (i.e., it is a burst
oscillation\cite{watts2012}) and is not coming from a background
source present within the field of view of XRT. Similar results were
reported by ref.~65 and 66\nocite{patruno2013,riggio2013}.

We also searched for oscillations during the other observations
performed by Swift, and found a signal significant above a 3$\sigma$
confidence level during the observation starting on 30 March 2013
(Obs.~ID 00552369000), with an amplitude of $7.4\pm1.9\%$.

%%%%%%%%%%%%%%%%%%%%%%%%%%%%%%%%%%%%%%%%%%%%%%%%

\section{Chandra observations of {\igr}}
\label{sec:chandra}

The {\em Chandra} X-ray observatory observed {\igr} on 29 April 2013, 
starting from 00:24:14 for 53\,ks, via Director's
Discretionary Time. The observations were performed using the High
Resolution Camera\cite{zombeck1995} (HRC--S) in timing mode, which has
a $6''\times30''$ field of view with a 16-$\mu$s timing resolution, but
no spectral information. Data were reduced with the CIAO 4.5
software. We first checked the data for the presence of solar flares
and extracted a new observation-specific bad-pixel file. We then ran a
degap correction, and cleaned the image for the hot pixels.

The new Chandra image of the M28 globular cluster was compared to the
several observations of the field performed by Chandra during the past
decade (see Table~\,\ref{tab:chandra} for a complete list, and Fig.~3 of
the main body of the Letter for a comparison).  We identified one
source (source 23 from ref.~60) to be an order of
magnitude brighter than in many previous Chandra observations.  We
used the CIAO tools {\tt wavdetect} and {\tt celldetect} to infer a
good position for the transient source, which was
RA=$18^{h}\;24^{m}\;32.527^{s}$, Dec=$-24^{\circ}\;52'\;08.58''$, with
0.6$''$ uncertainty (at 1-$\sigma$ confidence level, inferred
from a 0.3$''$ and a 0.5$''$ statistical and pointing accuracy,
respectively). The position of this transient source is consistent
within a 3-$\sigma$ confidence level with the XMM--Newton (see
Sec.~\ref{sec:xmm}), Swift (see Sec.~\ref{sec:swift}) and ATCA
position (see Sec.~\ref{sec:atca} below) of {\igr} (see Fig.~3 of
the main body of the Letter). We therefore  identify it as the
accreting X-ray pulsar, {\igr}.

We have extracted the source events from a 2$''$ region around the
position of the source (and background spectra far from the globular
cluster). In this Chandra observation the source had a count rate of
0.0251(5) counts per second (see Tab.\,\ref{tab:chandra}), which
assuming the spectral shape that {\igr} had in the closest Swift XRT
observation (a power-law with $\Gamma=2.5\pm0.6$; derived summing all
Swift observations taken in the week starting from 21 April 2013),
leads to a 0.5--10keV observed flux
$\sim4\times10^{-13}$\,erg\,cm$^{-2}$\,s$^{-1}$.

We searched for a coherent signal at the spin period of the source in
the time series corrected for the spacecraft and orbital motion, and
converted to the Solar System Barycentre. No signal was detected at a
confidence level of 3$\sigma$, with an upper limit of 17$\%$ on the
pulse amplitude, derived following the prescription given by Vaughan
et al. [65]. Pulsations at an amplitude lower than
such a value were observed during most of the XMM-Newton observations
(see Figure~\ref{fig:corr}), when the source was brighter by three
orders of magnitude. The non-detection of pulsations does not rule out
that the X-ray emission of {\igr} was pulsed during the Chandra
observation, at a level similar to that previously seen.

\begin{figure}[t!]
\centering
\includegraphics[width=16cm]{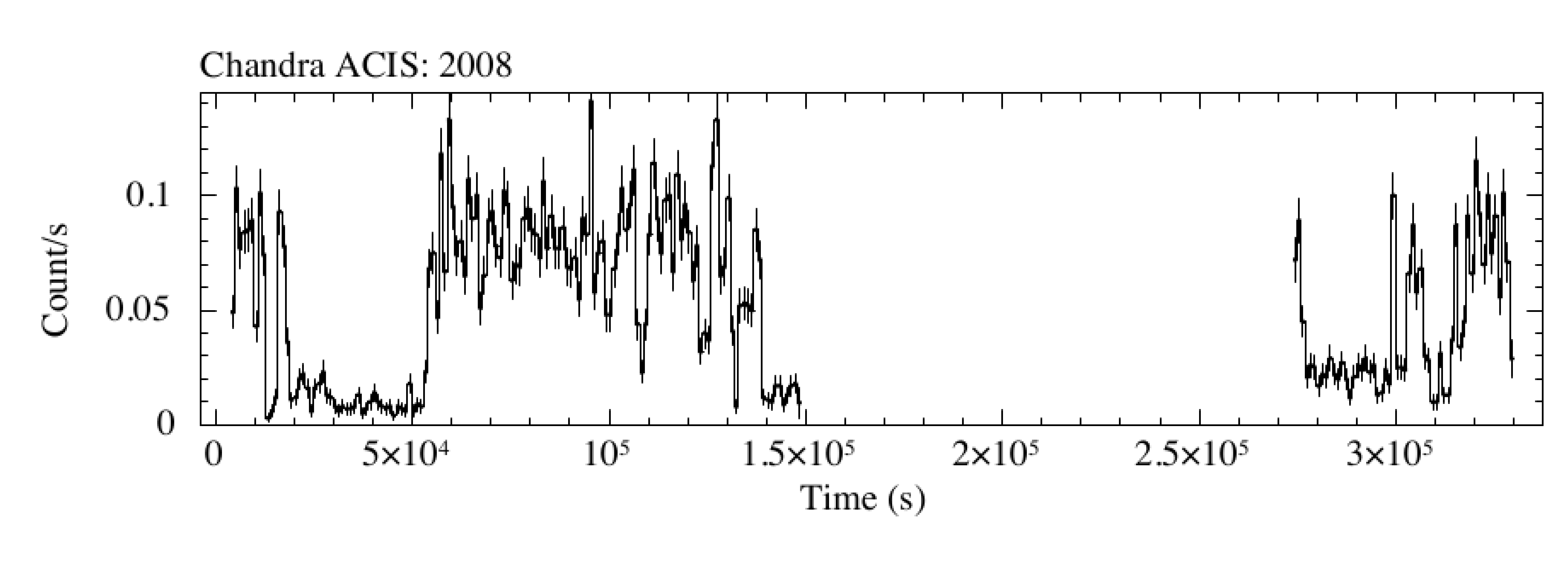}
\caption{ \footnotesize Chandra ACIS--S light curve of the August
2008 observations, when the source was undergoing an X-ray
outburst. Light curve is background subtracted, binned at 1000s and in
the 0.3--10\,keV energy range. Data were not available during the
interval between $1.5\times10^5$ and $2.7\times10^5$ s, since the
beginning of the first observation. \vspace{1cm}}
\label{fig:chlc}
\end{figure}

We have re-analysed all archival observations performed by Chandra in
the past decade to identify and follow the flux evolution of the
source. Timing and spectral data were always extracted from a 2$''$
region around the position of the source. A previous X-ray outburst of
{\igr} is detected during the August 2008 observations performed with
the Advanced CCD Imaging Spectrometer\cite{garmire2003} (ACIS; see
Fig.~3 of the main body of the Letter).  The light curve of this
previous outburst shows a strong variability on many timescales (see
 Figure~\ref{fig:chlc}). The most noticeable event is a total
quenching of the X-ray emission for a timescale of $\sim$10\,hrs,
compatible with a complete orbital period. We did not detect a
significant energy dependency of the light curve shape.

\begin{figure}[t!]
\centering
\includegraphics[width=12cm]{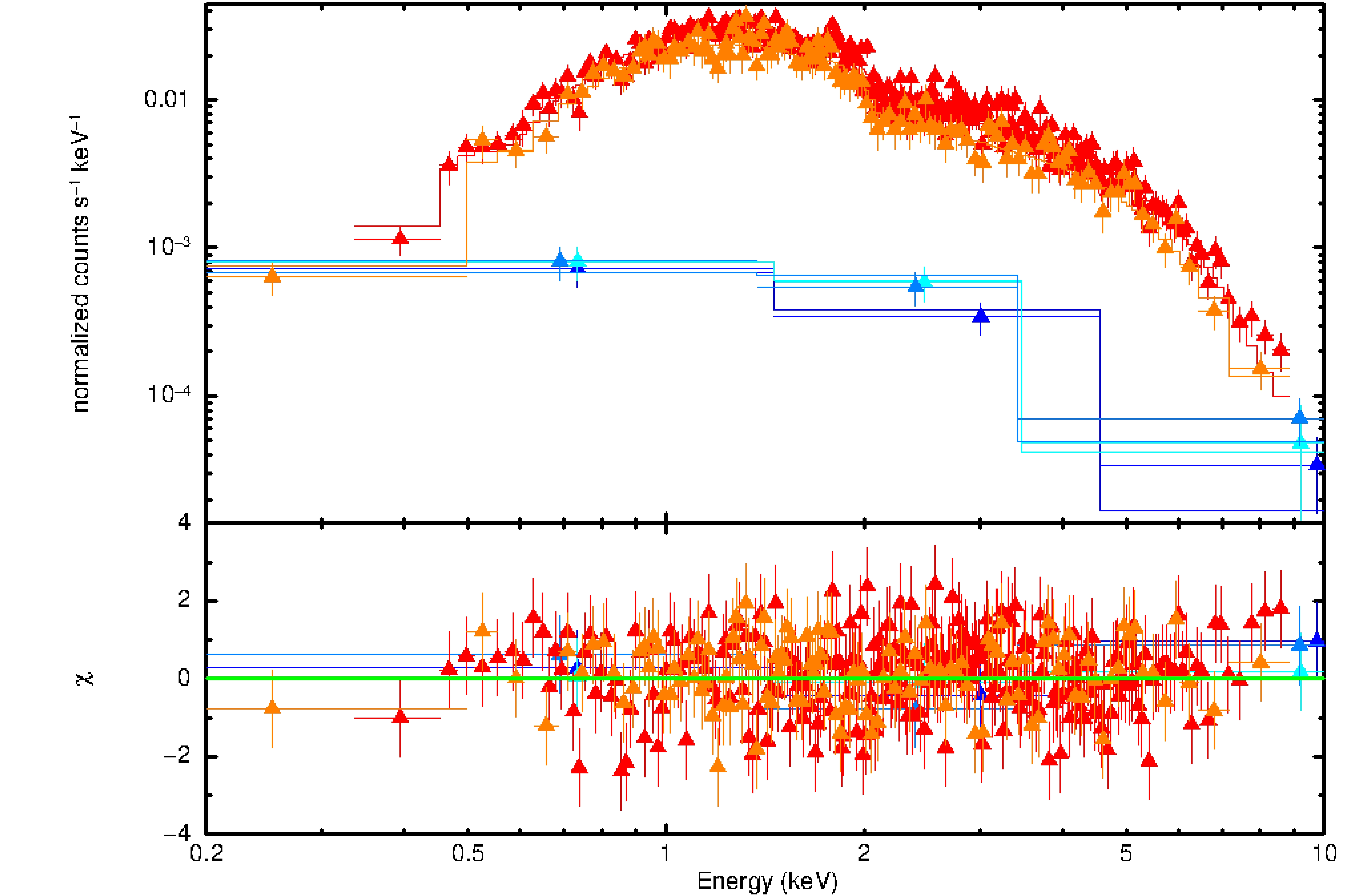}
\caption{ \footnotesize ACIS--S spectra of all {\em Chandra}
observations of {\igr} (4 July 2002, cyan triangles; 4 August 2002,
light blue triangles; 9 September 2002, blue triangles; 7 August 2008,
red triangles; 10 August 2008, orange triangles; see Supplementary
Table 2 for details) modelled with an absorbed power-law model (top
panel) and residuals with respect to the best fitting absorbed
power-law model (bottom panel).\vspace{1cm}}
\label{fig:chsp}
\end{figure}

Spectra were extracted from all the archival observations performed
with ACIS-S (all taken in VFAINT mode), which have a good spectral and
imaging resolution, but a 3.2\,s timing resolution, insufficient to
search for millisecond pulsations. We found a good fit
($\chi^2_{\nu}=0.91$ with 378 degrees of freedom) when modelling all
ACIS-S spectra together with an absorbed power-law model (leaving the
photoelectric absorption parameter, $N_{H}$, tied to be the same for
all spectra). We found a flux variability of more than one order of
magnitude between the 2002 and the 2008 observations of {\igr} (see
 Figure~\ref{fig:chsp} and Tab.\,\ref{tab:chandra}). In particular, we
found $N_{H}=0.26(2)\times10^{22}$\,cm$^{-2}$ (with abundances and
photoelectric cross-sections from ref.~69\nocite{anders1989} and 70\nocite{baluncinska1992}, respectively), a stable photon index
$\Gamma\sim1.5$, and a 0.5--10\,keV flux varying from 2.4 to
64$\times10^{-14}$\,erg\,cm$^{-2}$\,s$^{-1}$ (all errors are at 90$\%$
confidence level).

%%%%%%%%%%%%%%%%%%%%%%%%%%%%%%%%%%%%%%%%%%%%

\begin{table*}
\caption{{\em Chandra} observations of the Globular Cluster M28.}
\begin{center}
\begin{tabular}{@{}lcccccc}
\hline
Instrument & ObsID & Date & Exposure time (ks) & Source Count/s$^a$ &
$\Gamma^{b}$ & Flux$^c$\\
\hline
ACIS--S & 2684 & 2002-07-04 & 12.7 & 0.0029(6)& 1.5(0.7)& 3.1(9)\\
ACIS--S & 2685 & 2002-08-04 & 13.5 &  0.0030(5)& 1.4(0.7)& 3.5(9)\\
ACIS--S & 2683 & 2002-09-09 & 14.1 & 0.0025(4)& 1.6(0.7)& 2.4(7)\\
HRC--S & 2797 & 2002-11-08 & 48.3 & 0.0036(5)& 1.5 fix & 4.8 \\
HRC--S & 6769 & 2006-05-27 & 40.9 & 0.0036(5) & 1.5 fix & 4.8\\
ACIS--S & 9132 & 2008-08-07 & 142.3 & 0.0563(6) & 1.51(5)& 64(2)\\
ACIS--S & 9133  & 2008-08-10 & 54.5 & 0.04393(9) & 1.58(7) & 47(2)\\
HRC--S & 15645 & 2013-04-29 & 53.0 &  0.0503(8)& 2.5 fix & 41 \\
\hline
\hline
\end{tabular}
\end{center}
\begin{list}{}{}
\item[$^a$] We report the HRC--S counts converted in ACIS-S counts for
an easier comparison. The HRC--S original count rate is 0.0012(3) for
the first and second HRC--S observation, and 0.0251(5) for the more recent
one.
\item[$^b$] Absorption column density is
$N_{H}=0.26(2)\times10^{22}$\,cm$^{-2}$.
\item[$^c$] Absorbed, in the 0.5--10\,keV energy range, and in units
of $10^{-14}$\,erg\,cm$^{-2}$\,s$^{-1}$. HRC--S fluxes are estimated
assuming the source $N_{H}$ and a power-law spectrum with
$\Gamma=2.5$.

\end{list}

\label{tab:chandra}
\end{table*}

\section{ATCA observations of {\igr}}
\label{sec:atca}
On 2013 April 5 we observed {\igr} with the Australia Telescope
Compact Array (ATCA), at 9 and 5.5 GHz simultaneously. The data were
analysed with the Miriad package distributed by
ATNF\cite{sault1995}. Only one source was detected in the inner core
of the cluster, at a significance level of nearly 20$\sigma$ in both
frequencies, allowing for an accurate determination of its position at
RA=$18^{h}24^{m}32.51^{s}$ Dec=$-24^{\circ} 52' 07.9''$, with a 90\%
confidence error of 0.5$''$. This position is consistent with that
determined by Chandra (see Sec.~\ref{sec:chandra} and Fig.~3 of the
main body of the Letter).  The mean flux density of this source was
0.75$\pm$0.04 (0.62$\pm$0.03) mJy at 9 (5.5) GHz respectively,
yielding a mean spectral index of $0.4\pm0.2$. During the first 90
minutes of the observation, the source was strongly variable, reaching
up to 2.5 times the mean flux density.

The spectral properties of the ATCA source are similar to those
observed from other accreting millisecond pulsars\cite{gaensler1999},
and interpreted in terms of emission originating from shocks within
material ejected by the X-ray pulsar. Besides {\psr}={\igr}, which was
powered by accretion at that moment, none of the other
rotation-powered pulsars in the cluster were detected during the ATCA
observation.

\section{Parkes radio telescope observations of {\psr}}
\label{sec:parkes}

{\igr} was observed three times with the 64-m Parkes radio telescope
in 2013 April/May (see Supplementary Table \ref{tab:history}).  We
used the position determined by Chandra (see Sec.~\ref{sec:chandra}
above). The observations were carried out using simultaneously the
Berkeley-Parkes Swinburne Recorder (BPSR) and the Parkes Digital
Filterbank (PDFB4) in search mode, for the first epoch, and BPSR and
the analogue filterban (AFB), for the subsequent ones. The backends
were operating at central frequencies of 1382 MHz (BPSR) and 1369 MHz
(PDFB4 and AFB) , over bandwidths of 400 and 256 MHz respectively,
subdivided in 1024 frequency channels (512 for AFB).  The total usable
bandwidth for BPSR, after removal of a known interference from the
Thuraya3 satellite, is 350 MHz.

The resulting time series, 2-bit sampled every 64~$\mu$s for BPSR and
PDFB4, and 1-bit sampled every 80~$\mu$s for AFB, were folded with the
{\tt dspsr} package\cite{vanstraten2011} using the X-ray ephemeris
presented in this work and using a value of the dispersion
measure\cite{begin2006} of 119 pc cm$^{-3}$. {\psr} was detected in
two of three Parkes observations (see Supplementary Table
\ref{tab:history}).  In the case of the non-detection, the flux
density upper limit derived using the radiometer equation modified for
pulsars\cite{dewey1985} is $S_{min}=0.067$ mJy for a signal with a
signal-to-noise ratio SNR=8 and a pulse duty cycle of 15 per cent. We
caution that Parkes non-detection, as well as other radio
non-detections presented below, cannot be taken as strong evidence
that the radio pulsar was not active. {\psr} is well known to eclipse
in the radio\cite{bogdanov2011}, particularly around superior
conjunction, and this is likely the cause of at least some of the
radio non-detections presented in Supplementary Table
\ref{tab:history}.  Though there is no published flux density for
    {\psr}, given its comparable brightness to other pulsars in M28,
    it is likely that if the source was emitting as a radio pulsar
    during these observations that it would be at the threshold of
    detectability during this 64-m Parkes radio telescope observation
    - especially if the signal was further perturbed by intra-binary
    material, as in similar systems. Therefore, we consider these
    observations only moderately constraining as to whether the source
    was emitting as a radio pulsar at this epoch.

\section{Westerbork Synthesis radio telescope observations of {\psr}}

We observed {\igr} during four sessions in 2013 May using the
Westerbork Synthesis radio telescope (WSRT). The pointing position was
$18^{h}\,24^{m}\,32.496^{s}$, $24^{\circ}\,52'07.799''$.  We used WSRT
in the tied-array mode (gain, G = 1.2\,K/Jy), combining 13 of the
individual 25-m dishes in phase and recording with the PuMaII pulsar
data recorder\cite{karuppusamy2008}. We recorded a 160-MHz bandwidth
and coherently de-dispersed and folded the data offline using the {\tt
  dspsr} package\cite{vanstraten2011} and the X-ray-derived ephemeris
presented here.  {\psr} was just barely detected in one of four WSRT
observations (see Supplementary Table \ref{tab:history}).  Using the
radiometer equation modified for pulsar signals\cite{dewey1985}, we
can place a flux density limit of $S_{1400} = 0.08$ mJy on radio
emission from the pulsar in the case of the non-detections.  Note that
the flux of {\psr} is very close to WSRT's detection threshold and, as
mentioned above, the radio pulsar may simply have been eclipsed in the
case of non-detections.

\section{Green Bank Telescope observations of {\psr}}

 We observed {\igr} during seven sessions in 2013 May using the Green
 Bank Telescope (GBT).  The Chandra position presented here was used
 for pointing, and data were acquired using the GUPPI data
 recorder\cite{duplain2008} in a 800-MHz band centered at 2.0GHz.
 {\psr} was detected in two of seven observations (see Supplementary
 Table \ref{tab:history}).  Flux density limits in the case of
 non-detections are $S_{2000} \simlt 20\mu$Jy, but come with the same
 caveats about eclipsing as described above.  Supplementary Table
 \ref{tab:history} also lists a number of archival detections of
     {\psr} with the GBT since its discovery\cite{begin2006} as a
     radio pulsar in 2006.  These observations have been acquired as
     part of a regular timing program of the radio pulsars in M28
     (Ransom et al.) and together with archival X-ray observations
     they conclusively demonstrate that the system has switched
     between rotation-powered radio pulsar and accretion-powered X-ray
     pulsar (and back) during 2006--2013.

\LTcapwidth=\textwidth
\begin{footnotesize}
\begin{landscape}
%\begin{center}
\begin{longtable}{lclccc}
\caption[Abridged Radio/X-ray History of
  {\psr}/{igr}]{{\bf Abridged Radio/X-ray
    History of {\psr}/{\igr}}: Flux densities of
  radio observations were measured at 2.0 GHz (GBT), 1.4 GHz
  (WSRT/Parkes), and 5.5/9 GHz (ATCA). Radio non detections are marked
  by a dash in the flux column, and are only moderately constraining
  as to whether the source was emitting as a radio pulsar at this
  epoch (see text for details). GBT non-detections prior to 2012-10-07 are
    not listed. Observed X-ray fluxes are evaluated
  in the 0.5--10 keV  band (20--100 keV for INTEGRAL), and given
  in mCrab ($1$Crab = $4.32\times10^{-8}$ erg cm$^{-2}$ s$^{-1}$ in the
  0.5--10 keV band).} \label{tab:history} \\

\hline \multicolumn{1}{c}{UT Date} & \multicolumn{1}{c}{MJD Start} & \multicolumn{1}{c}{Telescope} & \multicolumn{1}{c}{Type}&  \multicolumn{1}{c}{Flux}&    \multicolumn{1}{c}{Comments}\\ \hline 
\endfirsthead

\multicolumn{6}{c}%
{{\bfseries \tablename\ \thetable{} -- continued from previous page}} \\
\hline \multicolumn{1}{c}{UT Date} & \multicolumn{1}{c}{MJD Start} & \multicolumn{1}{c}{Telescope} & \multicolumn{1}{c}{Type}&  \multicolumn{1}{c}{Flux}&    \multicolumn{1}{c}{Comments}\\ \hline 
\endhead

\hline \multicolumn{6}{|r|}{{Continued on next page}} \\ \hline
\endfoot

\hline \hline
\endlastfoot

2002-07-04   	   & 52,459.752	 & Chandra/ACIS--S		& X-rays	& $(7\pm2)\times10^{-4}$ mCrab	&		      X-ray Quiescence	\\
2002-08-04 	   & 52,490.990	 & Chandra/ACIS--S		& X-rays	& $(8\pm2)\times10^{-4}$	mCrab&		      	 X-ray Quiescence	\\
2002-09-09 	   & 52,526.705	 & Chandra/ACIS--S		& X-rays	& $(6\pm2)\times10^{-4}$	mCrab&		      	 X-ray Quiescence	\\
2002-11-08 	   & 52,586.237	 & Chandra/HRC--S		& X-rays	& $1\times10^{-3}$ mCrab	    	&		   	 X-ray Quiescence	\\
%2006-01-03         & 53,738	 & GBT		                & Radio	&				&			& Discovery of PSR~J1824$-$2452I (from Begin 2006)	\\
2006-01/02       & 53,738 -- 53,781      & GBT   & Radio	&				& Discovery of {\psr} (ref.~69)	\\
%-- 2006-02-15 & & & & & \\
2006-05-27 	   & 53,882.520	 & Chandra/HRC--S		& X-rays	& $1\times10^{-3}$ mCrab	    	&			X-ray Quiescence	\\
2006/2007 & & \multicolumn{3}{l}{Various GBT Radio Observations} \\
2007-12-30 	   & 54,464.746	& GBT		& Radio & $\sim 20\,\mu$Jy 	        	        & Radio Pulsations 	\\		
2008-04-17	   & 54,573.354	& GBT		& Radio	& $\sim 20\,\mu$Jy 	                 & Radio Pulsations 	\\
2008-06-13	   & 54,631.274	& GBT		& Radio	& $\sim 20\,\mu$Jy                          & Radio Pulsations 	\\
2008-08-07	   & 54,685.865	& Chandra	& X-rays	& $(14.8\pm0.5)\times10^{-3}$  mCrab   &			 X-ray Enhanced    	\\
2008-08-10 	   & 54,688.993 & Chandra	& X-rays	& $(10.9\pm0.5)\times10^{-3}$  mCrab  &			 X-ray Enhanced    	\\		      		
2009-05-06	   & 54,957.418	& GBT		& Radio	& $\sim 20\,\mu$Jy                        & Radio Pulsations 	\\
2009/2012 & & \multicolumn{4}{l}{Various GBT Radio Observations} \\
2012-10-07	   & 56,207.967		& GBT		& Radio	& --							&     --			\\
2013-01-06	   & 56,298.702			& GBT		& Radio	& --				&			      --			\\
2013-03-28   	   & 56,379.122 & INTEGRAL/ISGRI& X-rays	& 	 $\sim90$   mCrab			&			 Discovery of {\igr} \\
2013-03-30	   & 56,381.632	& Swift/XRT	& X-rays	& $(19.4\pm0.2)$	mCrab			&			 X-ray Outburst / Pulsations	\\	
2013-04-04	   & 56,386.018	& XMM-Newton/EPIC pn	& X-rays	& $(6.56\pm0.02)$ mCrab		&			X-ray Outburst / Pulsations	\\
2013-04-05	   & 56,387.720			& ATCA		& Radio	& $(0.62\pm0.03)$ mJy (5.5 GHz)	&			 Non-pulsed  \\
			   &			& 			&		& $(0.75\pm0.04)$ mJy (9 GHz)		& Non-pulsed	\\
2013-04-08	   & 56,390.481	& GBT		& Radio	& --				& 	--			\\
2013-04-13	   & 56,395.294	& XMM-Newton/EPIC pn	& X-rays	& $(6.76\pm0.02)$	mCrab			&		 X-ray Outburst / Pulsations	\\
2013-04-15	   & 56,397.469 & GBT		& Radio	& --				& 	--			\\
2013-04-29 	   & 56,411.010	& Chandra/HRC--S& X-rays$^{\,a}$	& $9.5\times10^{-3}$ 	 mCrab   			& X-ray Enhanced    	\\ 
		      		
2013-04-29	   & 56,411.560	& Parkes	& Radio	& --				&  	--			\\

2013-05-01         & 56,413.557 & Swift/XRT     & X-rays$^{\,a,b}$ & $(3.8\pm1.0)\times10^{-2}$ mCrab &  Latest X-ray detection \\

2013-05-02	   & 56,414.164	& WSRT		& Radio	& $50 \pm 30 \mu$Jy &  Radio Pulsations 	\\
2013-05-04	   & 56,416.162 & WSRT		& Radio	& --				& 	--			\\
2013-05-06	   & 56,418.170	& WSRT		& Radio	& --				& 	--			\\
2013-05-06	   & 56,418.296	& GBT		& Radio	& --				& 	--			\\
2013-05-06         & 56,418.831 & Swift/XRT     & X-rays$^{\,a,b}$ & $<4\times10^{-2}$ mCrab &  X-ray non detection \\
2013-05-07	   & 56,419.114	& WSRT		& Radio	& --				& 	--			\\
2013-05-09	   & 56,421.442	& GBT		& Radio	& --				& 	--			\\
2013-05-10	   & 56,422.544	& Parkes	& Radio	& $60 \pm 30 \mu$Jy  & Radio Pulsations	\\
2013-05-11	   & 56,423.405	& GBT		& Radio	& $10 \pm 5 \mu$Jy   & Radio Pulsations	\\
2013-05-13	   & 56,425.432	& GBT		& Radio	& $20 \pm 10 \mu$Jy  & Radio Pulsations	\\
2013-05-13	   & 56,425.684	& Parkes	& Radio	& $50 \pm 30 \mu$Jy  & Radio Pulsations	\\
2013-05-18	   & 56,430.277	& GBT		& Radio	& --			&	--			\\
2013-05-24	   & 56,436.437	& GBT		& Radio	& --			&	--			\\	
2013-05-31	   & 56,443.183	& GBT		& Radio	& --			&	--			\\

\end{longtable}

$^{\,a}$ Flux estimated assuming a $\Gamma=2.5$ power-law spectral
shape. $^{\,b}$ Other unresolved sources\cite{becker2003,bogdanov2011} in M28 are
expected to give a contribution of $\approx 7\times10^{-3}$ mCrab to
the quoted flux value/upper limit. Upper limits are evaluated at
3-$\sigma$ confidence level.
%\end{center}
\end{landscape}
\end{footnotesize}

\bibliography{nature.bib}

\end{document}